\documentstyle[12pt,worldsci,epsf]{article}

\def\intablecenterline#1{\hfill#1\hfill} 
\def\und{} 
\def\ETTunderline#1{\ifmmode\let\und=\underline\else 
\let\und=\underbar\fi\und{#1}}

\def\mathline{\ifmmode \vert\else$\vert$\fi}

\def\ETTstretch#1{\message{[[Sorry, you can't use Stretch with ETT]]}} 
 
\def\ETTexpand#1{\message{[[Sorry, you can't use Expand with ETT]]}} 
 
\def\xbackslash{\ifmmode \backslash\else$\backslash$\fi}

\def\quoteswitch{} 
\def\leftquotemarks{\ifmmode{}{}''\else``%
\let\quoteswitch=\rightquotemarks\fi} 
\def\rightquotemarks{\ifmmode{}{}''\else"\let\quoteswitch=\leftquotemarks\fi} 
\let\quoteswitch=\leftquotemarks 
\def\p$&-\${\overline{p}} 
\def\ETTstack#1#2{\ifmmode\ifx\int#1\displaystyle#1\limits#2\else 
\ifx\xint#1\displaystyle\int\limits#2\else 
\ifx\sum#1\displaystyle#1\limits#2\else%
\ifx\xsum#1\displaystyle\sum\limits#2\else 
\mathop{#1}\limits#2\fi\fi\fi\fi\else#1\mathop{#2}\limits\fi} 
 
\def\ETThalign#1{{\let\centerline=\intablecenterline 
\ifmmode\vcenter{\everycr={\noalign{\vskip1.5pt}}%
\halign{$\strut##$&$\strut##$&$\strut##$&$\strut##$&$\strut##$&%
$\strut##$&$\strut##$&$\strut##$&$\strut##$&$\strut##$&$\strut##$&%
$\strut##$&$\strut##$&$\strut##$&$\strut##$&$\strut##$&$\strut##$&%
$\strut##$&$\strut##$&$\strut##$\cr#1\crcr}\everycr={}}\else%
\hbox{$\vcenter{\everycr={\noalign{\vskip1.5pt}} 
\halign{\strut##&\strut##&\strut##&\strut##&%
\strut##&\strut##&\strut##&\strut##&\strut##&\strut##&\strut##&%
\strut##&\strut##&\strut##&\strut##&\strut##&\strut##&\strut##&%
\strut##&\strut##\cr#1\crcr}\everycr={}}$}\fi}} 
 
\gdef\ETTcolumn#1{\ifmmode%
\hbox{$\vcenter{\everycr{\noalign{\vskip-2pt}}%
\let\centerline=\intablecenterline 
\halign{\strut\hfill$##$\hfill\cr 
#1\crcr}\vskip-2pt}$}\else 
\hbox{$\vcenter{\offinterlineskip%
\let\centerline=\intablecenterline 
\halign{\strut\hfill##\hfill\cr 
#1\crcr}}$}\fi} 
 
\def\ETTtable#1{\ifmmode\vcenter{\everycr={\noalign{\vskip1.5pt}}%
\let\centerline=\intablecenterline%
\halign{\hfil$\vcenter{\hbox{$\strut##$}}$\hfil\ &\ %
\hfil$\vcenter{\hbox{$\strut##$}}$\hfil\ &\ %
\hfil$\vcenter{\hbox{$\strut##$}}$\hfil\ &\ %
\hfil$\vcenter{\hbox{$\strut##$}}$\hfil\ &\ %
\hfil$\vcenter{\hbox{$\strut##$}}$\hfil\ &\ %
\hfil$\vcenter{\hbox{$\strut##$}}$\hfil\ &\ %
\hfil$\vcenter{\hbox{$\strut##$}}$\hfil\ &\ %
\hfil$\vcenter{\hbox{$\strut##$}}$\hfil\ &\ %
\hfil$\vcenter{\hbox{$\strut##$}}$\hfil\ \cr 
#1\crcr}\everycr={}}\vrule depth3pt width0pt\relax\else 
\hbox{$\vcenter{\everycr={\noalign{\vskip1.5pt}}%
\let\centerline=\intablecenterline%
\halign{\hfil$\vcenter{\hbox{\strut##}}$\hfil\ &\ %
\hfil$\vcenter{\hbox{\strut##}}$\hfil\ &\ %
\hfil$\vcenter{\hbox{\strut##}}$\hfil\ &\ %
\hfil$\vcenter{\hbox{\strut##}}$\hfil\ &\ %
\hfil$\vcenter{\hbox{\strut##}}$\hfil\ &\ %
\hfil$\vcenter{\hbox{\strut##}}$\hfil\ &\ %
\hfil$\vcenter{\hbox{\strut##}}$\hfil\ &\ %
\hfil$\vcenter{\hbox{\strut##}}$\hfil\ &\ %
\hfil$\vcenter{\hbox{\strut##}}$\hfil\ \cr 
#1\crcr}\everycr={}}$}\vrule depth3pt width0pt\relax\fi} 
 
\def\ETTast{{\raise.25ex\hbox{$\ast$}}} 
 
\newif\ifinmath 
\newbox\superbox 
\newbox\superboxtwo 
 
\def\changetomath{$} 
 
\def\ETTsuperimpose#1#2{\ifmmode\let\check=\changetomath%
\else\let\check=\relax\fi%
\setbox\superbox=\hbox{\check#1\check}%
\setbox\superboxtwo=\hbox{\check#2\check}%
\ifdim\wd\superbox>\wd\superboxtwo%
\copy\superbox\hskip-\wd\superbox\hbox to 
\wd\superbox{\hfill\check#2\check\hfill}%
\else%
\copy\superboxtwo\hskip-\wd\superboxtwo\hbox to 
\wd\superboxtwo{\hfill\check#1\check\hfill}\fi} 
 
\def\hb{{\hskip3pt}} 
\let~=\hb

\newcount\moveup \newcount\moveover 
 
\def\lookagain{\futurelet\next\parser} 
 
\def\parser#1{\def\endb{}\ifx\next /\let\go=\relax \else\ifx\next r 
\global\advance\moveover by22\else 
\ifx\next l\global\advance\moveover by-22 
\else \ifx\next u\global\advance\moveup by22 
\else \ifx\next d\global\advance\moveup by-22\fi\fi\fi\fi 
\let\go=\lookagain\fi\go} 
 
\def\ETTadjust#1#2{\moveover=0 \moveup=0\setbox0=\hbox{\lookagain#1/}%
\divide\moveover by10 \divide\moveup by10\setbox1=\hbox{#2}%
\vtop to0pt{\hbox to\wd1{\hskip\moveover pt\raise\moveup pt\hbox{#2}\hss}%
\vss}} 
 
\newcount\upcounter 
\newcount\overcounter 
 
\newif\ifaz \newif\ifrz \newif\iflz \newif\ifdz \newif\ifuz 
 
\def\nonglobalreset{\azfalse\lzfalse\uzfalse%
\dzfalse\rzfalse}%
\long\gdef\ETTbox #1 #2{\def\parse{#1}%
\nonglobalreset\expandafter\looker\parse{}{}{}\boxer{\vbox{%
\let\centerline=\intablecenterline\halign{%
\strut##\cr#2\crcr}}}} 
 
\long\def\boxer#1{\ifaz\rztrue\lztrue\uztrue\dztrue\fi%
\hbox{$\,\,\vcenter{\ifuz\hrule\fi\hbox{\iflz\vrule\fi%
\hskip2pt\vbox{\vskip2pt\vbox{#1}\vskip2pt}%
\hskip2pt\ifrz\vrule\fi}\ifdz\hrule\fi}\,\,$}} 
 
\def\looker #1#2#3#4{%
\ifx#1A\aztrue\else\ifx#2a\aztrue\else
\ifx#1R\rztrue\else 
\ifx#2R\rztrue\else 
\ifx#3R\rztrue\else 
\ifx#4R\rztrue\fi\fi\fi\fi\fi 
\ifx#1L\lztrue\else 
\ifx#2L\lztrue\else 
\ifx#3L\lztrue\else 
\ifx#4L\lztrue\fi\fi\fi\fi 
\ifx#1U\uztrue\else 
\ifx#2U\uztrue\else 
\ifx#3U\uztrue\else 
\ifx#4U\uztrue\fi\fi\fi\fi 
\ifx#1D\dztrue\else 
\ifx#2D\dztrue\else 
\ifx#3D\dztrue\else 
\ifx#4D\dztrue\fi\fi\fi\fi\fi} 
 
\def\tie{\ifmmode\sim\else 
\hbox{\vtop to0pt{\hbox{\lower5pt\hbox{\~{}}}\vss}}\fi} 
\def\<{\ifmmode <\else $<$\fi} 
\def\>{\ifmmode >\else $>$\fi} 
 
\def\caret{\ifmmode{\vtop to0pt{\hbox{\lower7pt\hbox{$\hat{\vphantom{.}}$}} 
\vss}}\else \^{}\fi} 
 
\def\lcurlybracket{\ifmmode\{\else $\{$\fi} 
\def\rcurlybracket{\ifmmode\}\else $\}$\fi} 
 
\newdimen\tempdimen 
 
\def\smdarkP{\ \hbox{\global\setbox0=\hbox{\smallbackp}\vrule height\ht0 
depth\dp0 width0pt\vrule\hskip.6pt\vrule}\tempdimen=\ht0%
\advance\tempdimen by-.4pt%
\hskip-1pt\raise1.4pt\hbox{$\scriptstyle\bullet$}%
\hskip-2pt\llap{\raise\tempdimen\hbox to3pt{\hrulefill}}\ } 
 
\def\xxdarkP{\ \hbox{\global\setbox0=\hbox{\P}\vrule height\ht0 
depth\dp0 width0pt\vrule\hskip.6pt\vrule}\tempdimen=\ht0%
\advance\tempdimen by-.4pt\hskip-1pt\raise3pt\hbox{$\displaystyle\bullet$}%
\hskip-2pt\llap{\raise\tempdimen\hbox to4pt{\hrulefill}}\ }

\def\smallbackp{{\smallsymbol\char'173}}

\def\return{\hbox{\ \unskip{$\bf\leftarrow$\hskip-.5pt\raise2.4pt%
\hbox{\vrule height 2.5pt}}}}

\def\xxTM{\raise1ex\hbox{\amrseven TM}} 
 
\def\TMfive{\raise1ex\hbox{\amrfive TM}}

\def\xxeqcirc{\buildrel \lower1.5pt\hbox{$\scriptstyle\circ$}\over =} 
\def\smeqcirc{\buildrel \lower1.5pt\hbox{$\scriptscriptstyle\circ$}\over{\scriptstyle =}}

\def\xxeqdkcirc{\buildrel \lower1.5pt\hbox{$\scriptscriptstyle\bullet$}\over =} 
\def\smeqdkcirc{\buildrel \lower1.5pt\hbox{$\scriptscriptstyle\bullet$}\over{\scriptstyle =}}

\def\xxtridots{\ \unskip\raise4pt\hbox 
to1em{\hfil.\hfil}\llap{\hbox to1em{\hfil.\ .\hfil}}} 
 
\def\smtridots{\ \unskip\raise3pt\hbox 
to1em{\hfil.\hfil}\llap{\hbox to1em{\hfil.$\,$.\hfil}}}

\def\sqr#1#2{{\vcenter{\hrule height.#2pt 
\hbox{\vrule width.#2pt height#1pt\kern#1pt 
\vrule width.#2pt} 
\hrule height.2pt}}}

\def\xint{{\mathchoice{\int}{\displaystyle\int}{\int}{\int}}} 
\def\xsum{{\mathchoice{\sum}{\displaystyle\sum}{\sum}{\sum}}} 
 
\def\GermanS{\ifmmode\hbox{\ss}\else\ss\fi} 
\def\primeaccent{\ifmmode\hbox{\rm\char19}\else{\rm\char19}\fi} 
\def\underaccent{\ifmmode\hbox{\rm\char24}\else{\rm\char24}\fi} 
\def\EnglishPound{\ifmmode\hbox{\it\$}\,\else{\it\$}\fi} 
 
\newcommand{\be}{\begin{equation}}
\newcommand{\ee}{\end{equation}}
\newcommand{\bea}{\begin{eqnarray}}
\newcommand{\eea}{\end{eqnarray}}
\newcommand{\beas}{\begin{eqnarray*}}
\newcommand{\eeas}{\end{eqnarray*}}
\newcommand{\bi}{\begin{itemize}}
\newcommand{\ei}{\end{itemize}}
\newcommand{\bn}{\begin{enumerate}}
\newcommand{\en}{\end{enumerate}}

\def\lambar{{ \lambda \mkern-10mu\raise.5ex\hbox{--} }}

\def\thus{{ .. \mkern-7.5mu\raise.9ex\hbox{.} }\  }

\def\ba2#1#2{${\overline{#1}}^{#2}$}
\def\anti#1#2{\vbox{\ialign{##\crcr
     \hrulefill$\smash{\phantom{\scriptstyle#2}}$\crcr
     \noalign{\kern-1pt\nointerlineskip\vskip 0.25ex}
     $\hfil{#1}^{#2}\hfil$\crcr}}}
\def\anth#1#2{\vbox{\ialign{##\crcr
    \hrulefill$\smash{\phantom{\scriptstyle#2}}$\crcr
    \noalign{\kern-0.5pt\nointerlineskip\vskip 0.25ex}
    $\hfil{#1}^{#2}\hfil$\crcr}}}

\def\stack{\stackrel}

\begin{document}
\rightline{JLAB-THY-98-35}
\rightline{WM-98-104}
\title{CONFINEMENT THROUGH A RELATIVISTIC GENERALIZATION OF THE LINEAR
INTERACTION}
\author{\c{C}etin \c{S}avkl{\i}$^1$ and Franz Gross$^{1,2}$\\
\vspace{0.3cm}
{\em $^1$Department of Physics, College of William and Mary, Williamsburg,
Virginia 23187, USA\\
$^2$Jefferson Lab, Newport News, Virginia 23606, USA}}
\maketitle
\setlength{\baselineskip}{2.6ex}
\vspace{0.7cm}

\begin{abstract}

Generalizing a covariant framework previously developed, it is shown that
confinement insures that meson $\rightarrow q+\bar{q}$ decay amplitudes
vanish when both quarks are on-shell.  Regularization of
singularities in a  covariant linear potential associated with
nonzero energy transfers (i.e. $q^2=0, q^{\mu}\neq0$) is improved.
\end{abstract}
\vspace{0.5cm}

\section{Introduction}

Even for a simple system such as a  quark-antiquark (referred
to collectively as ``quarks'') bound state, color confinement implies that
the matrix element for meson decay, $\mu \rightarrow q + \bar{q}$, must
vanish whenever this decay is kinematically possible.   This trivial
statement can be realized by two possible  mechanisms.  Either {\bf i)} the
quark propagators are free of timelike mass poles, as is usually assumed in
Euclidean metric based studies~\cite{CETIN}, or {\bf ii)} the vertex
function of the bound state {\bf vanishes} when both quarks are on-shell.
In an earlier work~\cite{GROSS} it was assumed that the quark propagators
have mass poles, but it was not shown that relativistic generalization of
the linear confining potential guaranteed the correct vanishing of the vertex
function.  In  this work it is shown that the vanishing of the vertex
function is a general feature of any confining interaction, and that {\bf
insisting on the correct nonrelativistic limit leads naturally to the second
option.}

\section{The nonrelativistic linear potential}

It was shown earlier\cite{GROSS} that the nonrelativistic linear interaction
\begin{equation}
{\cal V}(r)=-C+\sigma r,
\end{equation}
can be expressed in momentum space by
\begin{equation}
{\cal V}(\vec{q})=\lim_{\epsilon\rightarrow
0}\Bigl[V_A(\vec{q})-\delta^3(\vec{q})\int
d^3q'V_A(\vec{q}')\Bigr]-(2\pi)^3\delta^3(\vec{q}) C,
\label{nrpotential}
\end{equation}
where
\begin{equation}
V_A(\vec{q})\equiv -8\pi\sigma\frac{1}{(\vec{q}^2+\epsilon^2)^2}\, .
\label{va}
\end{equation}
In coordinate space
\begin{equation}
V_A(\vec{r})=\int\frac{d^3q}{(2\pi)^2}\,e^{-iq\cdot r}\,V_A(\vec{q})=
-\sigma\frac{e^{-\epsilon r}}{\epsilon}\ \ \propto\ \
\lim_{\epsilon\rightarrow 0} \;\sigma (r-\frac{1}{\epsilon})\, ,
\end{equation}
and hence the delta function subtraction cancels the infinite $1/\epsilon$
term, insuring that the limit $\epsilon\to0$ exists and that the potential
confines.  It also insures that the linear potential vanishes at
the origin. In fact, the delta function subtraction could be used to
construct confining potentials for any monotonically increasing
$V_A(r)$ for which $V_A(r)-V_A(0)=\infty$ for some $r$. For example, after
the delta function subtraction even the coulomb potential
$V_A(r)=-{1}/{r}$ would result in a confining interaction.

Using the linear potential Eq.~(\ref{nrpotential}) the Schr\"odinger
equation for two quarks of equal mass $m$ becomes
\begin{equation}
\biggl[ \frac{\vec{p}^2}{m}-E\biggr]\Psi(\vec{p})=-\int
\frac{d^3k}{(2\pi)^3}
V_A(\vec{p}-\vec{k})\,\left[\Psi(\vec{k})-\Psi(\vec{p})\,\right]+
C\Psi(\vec{p})\, ,
\label{Schroedinger}
\end{equation}
where $E$ is the binding energy.

\section{The relativistic generalization}

A seemingly natural candidate for a relativistic generalization of the linear
interaction (\ref{nrpotential}) is to use the Bethe-Salpeter equation with
a kernel given by
\begin{equation}
{\cal V}(\vec{q})\stack{?}{\rightarrow}
\lim_{\epsilon\rightarrow 0}\left[V_A(q)-\delta^4(q)\int
d^4q'V_A(q')\,\right]-(2\pi)^3\delta^4(q) C\, ,
\end{equation}
where $\vec{q}^2\to-q^2$ in Eq.~(\ref{va}).  Unfortunately this
form does not have the correct nonrelativistic limit. Therefore
we rephrase the question: Can we find a covariant equation
that reduces to the Schr\"odinger equation with linear interaction in the
nonrelativistic limit? In order to motivate the relativistic equation,
start by  defining the Schr\"odinger {\em vertex} function $\Phi(\vec{p})$
\begin{equation}
\Phi(\vec{p})\equiv\left[
\frac{\vec{p}^2}{m}-E\right]\Psi(\vec{p})\, .\nonumber
\end{equation}
The Schr\"odinger equation with linear interaction is then
\begin{eqnarray}
\Phi(\vec{p})=-\int \frac{d^3k}{(2\pi)^3}\,m V_A(\vec{p}-\vec{k})
\left[\frac{\Phi(\vec{k})} { \left(\vec{k}^2-mE\right)}
-\frac{\Phi(\vec{p})}{ \left(\vec{p}^2-mE\right)}\right] +
mC\frac{\Phi(\vec{p})}{ \left(\vec{p}^2-mE\right)}\, .
\label{seq2}
\end{eqnarray}
Introducing the four-vectors $P$, $k$, and $p$, with $P=(2m+E,\vec{0})$ and
$p^2=k^2=m^2$ (so that $E_k=\sqrt{m^2+\vec{k}^2}$), the
relativistic generalization of (\ref{seq2}) which reduces to it in the
nonrelativistic limit (when $m\to\infty$) is
\begin{eqnarray}
\Phi(p)=&-&\int \frac{d^3k}{(2\pi)^3} \frac{2m^2}{E_k}
V_A(p-k)\,\left[\frac{\Phi(k)}{m^2-(P-k)^2}-\frac{\Phi(p)}{m^2-(P-p)^2}
\right]\nonumber\\
&+&2mC\frac{\Phi(p)}{m^2-(P-p)^2}\, . \label{geq}
\end{eqnarray}
%

\begin{figure}[thb]
\centerline{
   \epsfxsize=2.0in
   \epsfysize=0.5in
\epsffile{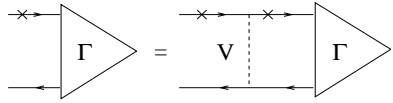}}
\caption{The Gross equation. The $\times$ indicate that the particle is
on-shell.}
\label{gross.fig}
\end{figure}

\noindent This is the Gross equation (see Fig.~\ref{gross.fig}) and the
discussion shows that this is the most natural relativistic equation
obtained from a generalization of the  nonrelativistic Schr\"odinger
equation for confined particles.

In a previous application of the Gross equation\cite{GROSS} the kernel
(\ref{va}), with $\vec{q}^2\to-q^2$, was used.  This choice is
undesirable for use with the two channel version of this equation, where the
mass shell constraints which fix $q_0$ introduce singularities for non-zero
momentum transfers (i.e. $q^2=0$, but $q^{\mu}\neq0$).  To correct this
problem the kernel $V_A(q)$ is defined to be
\begin{equation}
V_A(q)\equiv -8\pi \sigma\frac{1}{q^4+(P\cdot q)^4/P^4}
\label{nqtothe4}
\end{equation}
Advantages of this form are many: {\bf i)} singularities are restricted to
$q^{\mu}=0$, {\bf ii)} interaction strength {\em does not} depend on the bound
state momentum $P$ in the rest frame, {\bf iii)} it has the correct
nonrelativistic dependence on $\vec{q}^2$, and {\bf iv)} the ultraviolet
regularization used previously\cite{GROSS} is no longer needed.

\section{Proof of confinement}

\begin{figure}[thb]
\centerline{
   \epsfxsize=2.0in
   \epsfysize=0.5in
\epsffile{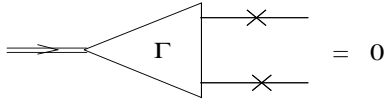}}
\caption{When both constituents are on-shell, the vertex function vanishes.}
\label{conf.fig}
\end{figure}

It is sufficient to consider the case when the constant term $C=0$.  If the
mass of the bound state $\mu>2m$, then there exists a value of three momentum
$|\vec{p}|=p_c$ when both quarks can be on-shell.  In this case
$(P-p)^2=\mu^2-2\mu E_{p_c}+m^2=m^2$, and the subtraction term in
Eq.~(\ref{geq}) appears to be singular.  This singularity is not cancelled
by the first term, and hence, if the equation is to have a solution, the
vertex function must be zero at $p_c$. In particular, as $|\vec{p}|\to
p_c\pm\epsilon$,
\begin{equation}
(P-p)^2-m^2\to \mp \frac{2\mu p_c\,\epsilon}{E_{p_c}}\, ,
\end{equation}
and the vertex function would have an infinite discontinuity at $p_c$
unless it were zero there.  We conclude that
$\Phi(p)=0$, when $2E_p=\mu$ (as illustrated in Fig.~\ref{conf.fig}).

\begin{figure}[thb]
\begin{center}
\mbox{
   \epsfysize=3.5in
\epsffile{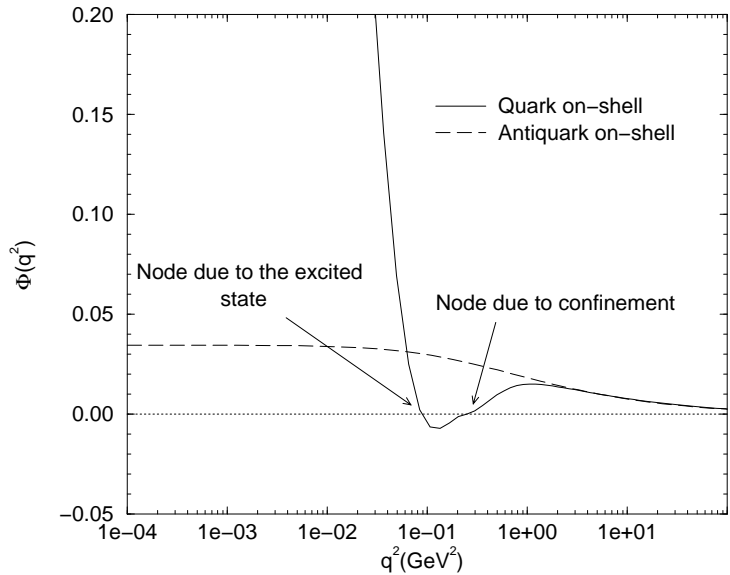}
}
\end{center}
\vskip -0.5in
\caption{The pion 1st excited state Gross vertex functions
are shown. The first node is due to the excited state.  The second node
assures that the bound state does not decay.}
\label{pistar.fig}
\end{figure}

A numerical confirmation of this result is shown in Fig.~\ref{pistar.fig}.
In the actual calculations to be presented in an upcoming
paper, the bound state equation is averaged over positive and negative
energy contributions and symmetrized by picking up the pole
contributions of both constituents.  The average over positive and negative
energy contributions leads to a two channel equation with a two
component vertex function.  In Fig.~\ref{pistar.fig} the solid line is the
(large) component in which the off-shell quark has positive energy and the
dashed line is the (small) component in which the off-shell quark has
negative energy. Since a physical decay must produce two positive energy
quarks, only the large component must have the ``confinement'' node.
Fig.~\ref{pistar.fig} shows the vertex functions for an excited pion with
mass $\mu=1.2$ GeV composed of two quarks with masses $m=0.34$ MeV.  The
large vertex function must have {\it two\/} nodes: one due to the
excitation and another at
$p^2=0.244$ GeV$^2$, exactly where both particles are
on-shell.

\vskip -0.3in

\section{Conclusions}
  The relativistic equation obtained from a generalization of the
nonrelativistic Schr\"odinger equation for confined particles is of the
Gross  equation type rather than the Bethe-Salpeter equation type.
The confinement mechanism arises not from the lack of quark
mass poles,  but through the vanishing of the vertex function when both
quarks are on their positive energy mass-shell. Because of this confinement
mechanism, the vertex  functions (ground or excited state) have one
additional node if the bound state is heavy enough.

\vspace*{0.3cm}

\section*{Acknowledgement}
This work was supported in part by the US Department
of Energy under grant No.~DE-FG02-97ER41032.

\end{document}